# Near-field focusing of dielectric microspheres: Super-resolution and field-invariant parameter scaling

Zengbo Wang[1], Yi Zhou[2], Boris Luk'yanchuk[2]

[1]School of Electronic Engineering, Bangor University, Dean Street, Bangor LL57 1UT, Gwynedd, UK
[2]Data Storage Institute, 5 Engineering Drive 1, Singapore 117608, Singapore.

Abstract:

Optical near-fields of small dielectric particles are of particular importance and interests for nanoscale optical engineering such as field localization, fabrication, characterization, sensing and imaging. This paper represents a systematic investigation on the focusing characteristics (focal length, field enhancement, spot size) for a given refractive-index microsphere (n=1.6) with a varying size parameter $q_0$ across the range of $\pi \leq q_0 \leq 20\pi$. Conditions for super-resolution foci were analysised in details. Particularly strong super-resolution foci with spot size falling at least 50% below the diffraction limit were identified and possible new applications were suggested. To understand how the super-resolution conditions could be scaled to other refractive-index particles or background medium, principles of field-invariant parameters scaling (size, wavelength, and refractive index) were revealed and demonstrated with example cases. It offers the new freedom to choose particles and background medium to gain super-resolution at any frequency across the whole electromagnetic spectrum.

# 1. Introduction:

The fields of nanophotonics and metamaterials have experienced an explosive growth in the past decade. Superlens [1], invisibility cloaking [2] and nanolasers [3] are possibly the best known examples. One general ingredient in these developments is the use of metals (e.g, gold, silver) that supports surface plasmon resonance [4]. The progress is currently limited by the intrinsic loss of metals. New strategies are constantly sought to tackle the loss issue. Examples include the use of gain medium to compensate the loss[5], and the adaption of high-index pure-dielectrics materials in the design [6]. This paper studies light interaction with pure-dielectric micro- and nanospheres, aiming to fully understand its near-field focusing properties which are important to super-resolution applications including field localization, fabrication[7, 8], characterization, sensing and imaging[9].

More than a decade ago Lu and Luk'yanchuk et al. first demonstrated that enhanced optical near-fields of a 500-nm silica sphere can be used for subwavelength structuring of silicon surface [10]. Since, there has been a strong continuing interest on the technique and many progresses have been made. Theoretical studies related to the topic also grew rapidly, as reviewed in [11]. Not knowing the initial work by Lu and Luk'yanchuk, four years later Chen et al. coined a new term 'photonic nanojet' for dielectric particle super-resolution focusing at the shadow side of the particle [12], which has since became popularly used in the field [13-20]. The reported super-resolution of the nanojet is on the scale of $\lambda/2n$ [21], where $\lambda$ is light wavelength and n the refractive index of particle. In case of a particle with $n = 1.6$, the resolution limit is about $\lambda/3.2 \approx 0.313\lambda$ following the literature. In this paper, we extend our investigation scope to include the internal particle focusing which haven't been well explored. Super-resolution limit as small as of $\lambda/5.6$ has been observed for some size parameter particles with $n = 1.6$. Such highly localized internal particle focusing is important for many non-linear optical applications such as high-order harmonic generation in optical materials. Another motivation of the present work is to look at how the parameters such as particle size, refractive index and background medium could be systematically scaled according to the practical conditions while attaining the near-fields invariant for super-resolution applications. This work offers the experimentalists new freedom to choose particles and background medium to gain super-resolution at any frequency across the whole electromagnetic spectrum. The paper is organized as follows. Section 2 describes the simulation methodology. Section 3 presents key results on super-resolution focusing characteristics including focal length, field enhancement and spot size analysis. Section 4 describes the principles of field-invariant scaling of Mie parameters. Section 5 concludes the paper.

# 2. Simulation Methodology:

The near-fields of a small sphere can be exactly calculated with classical Mie theory [22]. In case of non-magnetic particles, there are four general input parameters: incident wavelength of light ($\lambda$); medium refractive index ($n_m$); particle radius ($a$) and particle refractive index ($n_p$). These parameters can be reduced to two independent size parameters ($q_p, q_m$), defined as:

$$\text{Size parameter: } q_0 = \frac{2\pi a}{\lambda} \text{ ;}$$

(1)

$$\text{Particle size parameter: } q_p = q_0 . n_p = \frac{2\pi a}{\lambda} . n_p \text{ ;}$$

(2)

$$\text{Medium size parameter: } q_m = q_0 . n_m = \frac{2\pi a}{\lambda} . n_m$$

(3)

The near-field profiles are fundamentally determined by the scattering wave coefficients ($a_\ell, b_\ell$) and internal wave coefficients ($c_\ell, d_\ell$). For the convenience of parameter scaling discussion in following Section 4, both coefficients are represented here:

$$a_\ell = \frac{q_p \psi'_\ell(q_m) \psi_\ell(q_p) - q_m \psi_\ell(q_m) \psi'_\ell(q_p)}{q_p \varsigma'_\ell(q_m) \psi_\ell(q_p) - q_m \psi'_\ell(q_p) \varsigma_\ell(q_m)} ,$$

(4)

$$b_\ell = \frac{q_p \psi'_\ell(q_p) \psi_\ell(q_m) - q_m \psi_\ell(q_p) \psi'_\ell(q_m)}{q_p \psi'_\ell(q_p) \varsigma_\ell(q_m) - q_m \psi_\ell(q_p) \varsigma'_\ell(q_m)} ,$$

(5)

$$c_\ell = \frac{q_p \varsigma_\ell(q_m) \psi'_\ell(q_m) - q_p \varsigma'_\ell(q_m) \psi_\ell(q_m)}{q_p \varsigma'_\ell(q_m) \psi_\ell(q_p) - q_m \psi'_\ell(q_p) \varsigma_\ell(q_m)} ,$$

(6)

$$d_\ell = \frac{q_p \varsigma'_\ell(q_m) \psi_\ell(q_m) - q_p \varsigma_\ell(q_m) \psi'_\ell(q_m)}{q_p \psi'_\ell(q_p) \varsigma_\ell(q_m) - q_m \psi_\ell(q_p) \varsigma'_\ell(q_m)}$$

(7)

As will be shown below in section 4, ($q_p, q_m$) parameters can be equivalently expressed using ($q_0, n_p$) parameters, so that there is a freedom to chose using either ($q_p, q_m$) or ($q_0, n_p$). In this study, we will assign Mie parameter using ($q_0, n_p$). For completed Mie formulas please refer to [23]. The simulation program was coded in FORTRAN, which is available under request.

To simplify the calculation, following assumptions were made for the super-resolution focusing study in Section 3: $n_p$ =1.6 (fixed), $n_m$ =1.0 (fixed), size parameter $q_0$ varies from $\pi$ to $20\pi$ ( i.e., $\pi \leq q_0 \leq 20\pi$ ) with step resolution of 0.1. To find the particle focus, we calculate the $|E|^2$ values along a line across the particle centre from

-5a to 5a following the incident direction. Five thousand points were sampled in the calculation window, thus the result resolution is 500 points per particle radius. The focus size was measured as FWHM (Full Width at Half Maximum).

To compare our results with the diffraction limit, we can re-format the Rayleigh formula as the function of size parameter $q_0$, by

$$q_0 = \frac{2\pi a}{\lambda} \Rightarrow \lambda = \frac{2\pi a}{q_0} \tag{8}$$

$$\frac{D_{Rayleigh}}{a} = \frac{0.61\lambda}{a} = 0.61\frac{2\pi}{q_0} = \frac{1.22\pi}{q_0} \tag{9}$$

The diffraction limit curve following Eq. (9) can then be plotted on the same figure with our program outputs for a direct comparison of the calculated focus size with Rayleigh limit.

## 3. Near-fields of a dielectric sphere: focal length and super-resolution

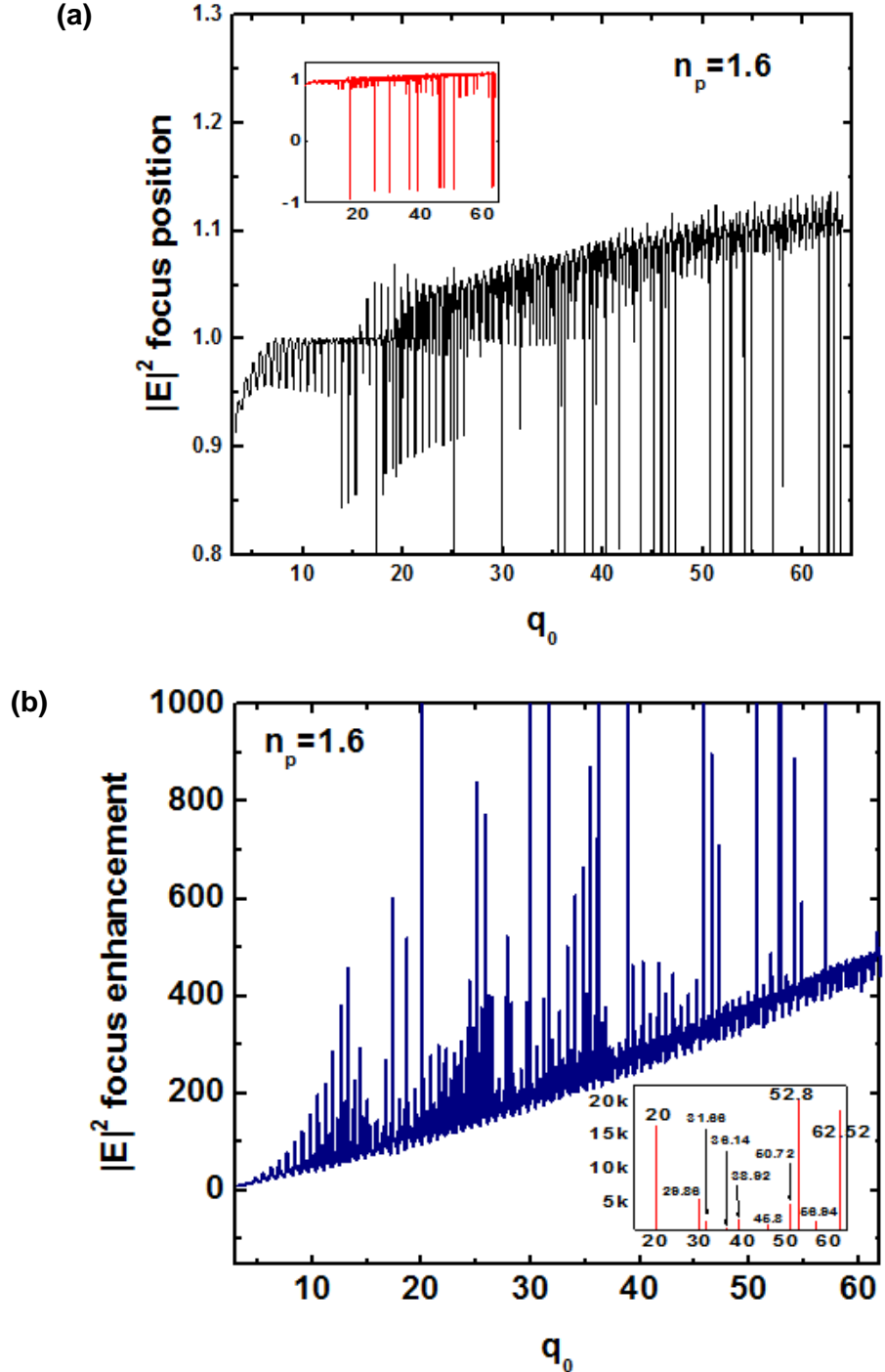

Figure 1: (a) Normalized focus position of $|E|^2$ field as a function of size parameter $q_0$. The inset shows the overall view including the negative focal length. (b) Corresponding field enhancement factor. Note y scales are different in both plots and insets.

Figure 1(a) shows the dependence of particle focus position of the $|E|^2$ field on varying size parameters. The position value was normalized with respect to the particle radius, which means that for position =1 the focus is at the bottom of particle while for position = -1 the focus is at the top of particle. The focus position for size parameters within the investigated range $\pi \leq q_0 \leq 20\pi$ shows an increasing tendency, but with highly oscillating values between 0.8 and 1.15. These oscillations are a typical result of the optical resonance within the spherical cavities. Such problem was discussed in Ref. [3]. For some size parameters, the focal length can be negative-valued which suggests focusing near the top side of particle (see inset) instead of the usual bottom side. It appears to be an indication of extraordinary sharp focusing as shown below. Figure 1(b) shows the corresponding focus enhancement factor as a function of size parameter. Similar to Fig. 1(a), the focus enhancement curve also manifests a general increasing tendency but with strong oscillations. Again, this is due to the high sensitivity of optical resonance inside the spherical cavities on the size parameter. It is surprising to see the giant field enhancements. Figure 1(b) inset shows the ten size parameters that produce field enhancement larger than 1k. Three of them, $q_0 = 20, 52.8, 62.52$, in fact produces enhancement larger than 10k- this is a surprising result in terms of its strength for dielectric particle focusing which would be of particular interests for those working in the field of nonlinear optics.

Figure 2 shows the $|E|^2$ near-field distributions for the special size parameters $q_0 = 20$ and $q_0 = 52.8$. In both cases, highly focused, nearly-symmetric foci are seen on both top and bottom sides of the particle. It is noted that the near fields at out-of-focus regions are weak and almost negligible as compared to the peak focus. Similar simulations were also performed for the other size parameters producing giant focus enhancement and confirmed its general similarity. The underlying physics for these foci relates to the singularities associated with internal wave coefficients ($c_\ell, d_\ell$) in Eqs. (6) and (7), and is still under investigation.

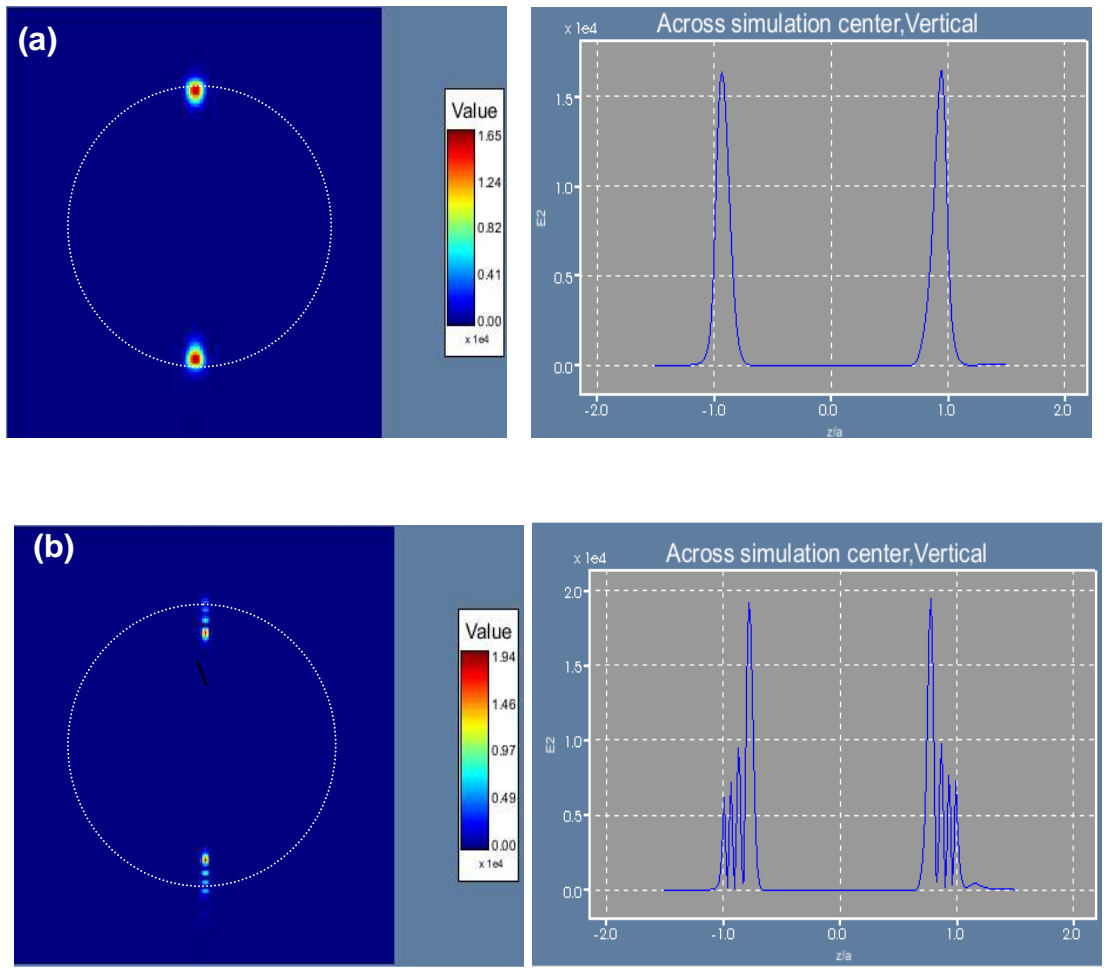

Figure 2 : $|E|^2$ field intensity distribution for (a) $q_0=20$ and (b) $q_0=52.8$ with $n_p=1.6$. Note that fields out of foci are almost zero and energy only concentrates on two nearly-symmetric foci points near top and bottom of the particle. The foci enhancement values are extremely strong, with enhancement of 16406 for $q_0=20$ and 20120 for $q_0=52.8$. $q_0=62.52$ has a similar field profile to $q_0=52.8$ and is not shown here.

Figure 3 shows calculated FWHM size in x- and y-direction within the cut-plane that across focus plane as in Fig.1. In Fig.4, the focus shape deviation factor D, defined by the difference of focus size along x and y direction was shown.

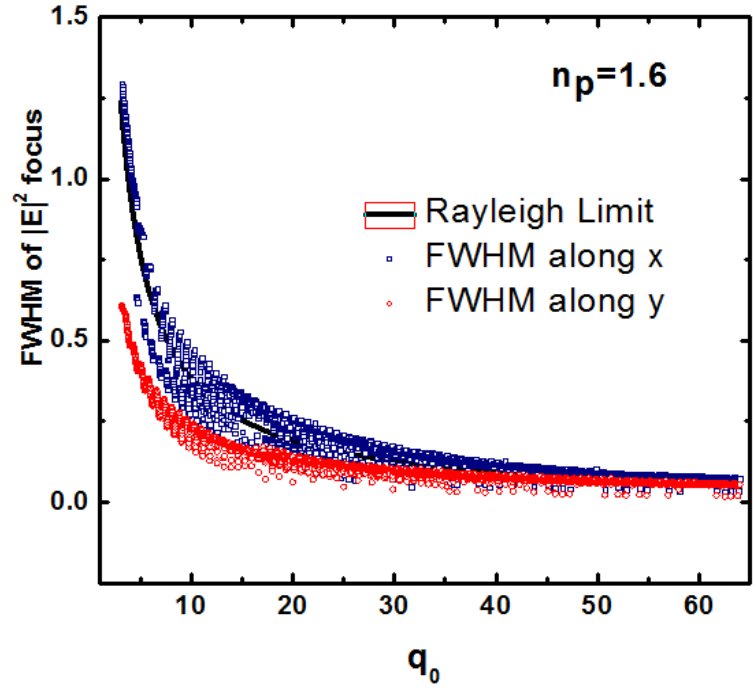

Figure 3: FWHM of $|E|^2$ focus as a function of size parameter $q_0$, the size of FWHM was normalized to particle radius.

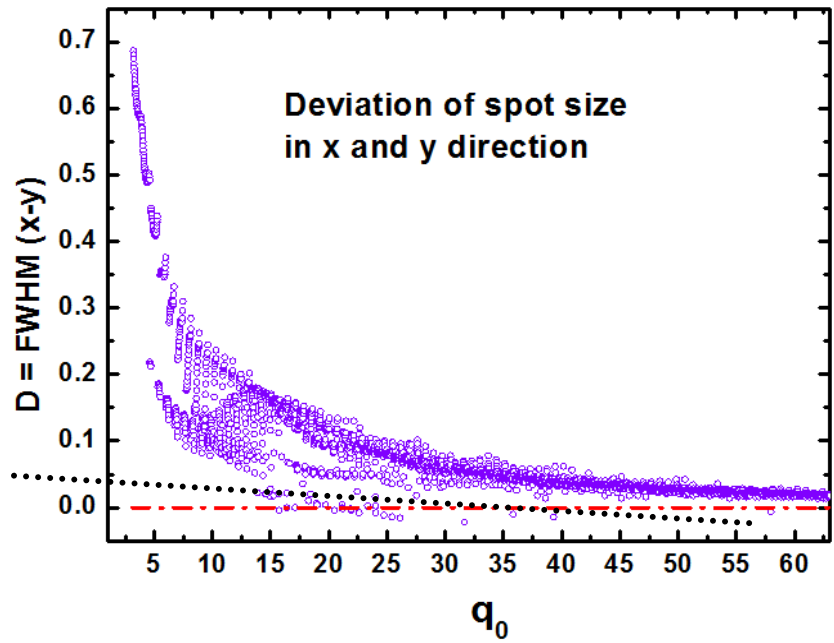

Figure 4: Difference of focus size along x-direction and y-direction (D value), which produces an elliptical focus.

From Figs. 3 and 4, one can see that the focus FWHM along y-direction (incident field polarized at x-direction) is always smaller than or very close to that along x-direction, meaning an elliptical spot profile. Meanwhile, the FWHM in y-direction is also generally smaller (i.e., y-direction super-resolution) than or close to the Rayleigh diffraction limit, and FWHM in x-direction manifests oscillating characteristics around the Rayleigh limit values. These findings indicate that the polarization property of the incident beam plays some role on the focus size of the electric field: the focus is prolonged at the polarization direction and shortened at cross direction.

A real super-resolution spot requires both values for FWHM in x and y directions below the Rayleigh limit curve (black curve in Fig.3). To find these super-resolutions spots, a new figure derived from Fig. 3 results was presented in Fig. 5. It shows the values of FWHM along x-direction, namely FWHM(x), minus Rayleigh limit and FHWM along y-direction, namely FWHM(y), minus Rayleigh limit. Negative values represent super-resolution in that direction. Real super-resolutions spots are seen for those size parameters that give both negative value in x and y direction. A lot of these super-resolution cases can be seen in Fig.5. Within them, we have indentified 22 'extreme' super-resolution spots whose average spot size are at least 50% smaller than the diffraction limit, as shown in Fig. 6 and tabulated in Table 1. The best super-resolution is seen for q=62.52, whose spot size in y-direction is 70.64% below the diffraction limit, corresponding to a record resolution of $0.179\lambda \approx \lambda/5.6$. Due to an elliptical shape nature of the focus, the resolution in x-direction is slightly bigger, but still below the diffraction limit by 47.8%. The averaged spot size for this case, calculated as square root of FWHM(x) * FWHM(y), is about $0.256\lambda \approx \lambda/3.9$.

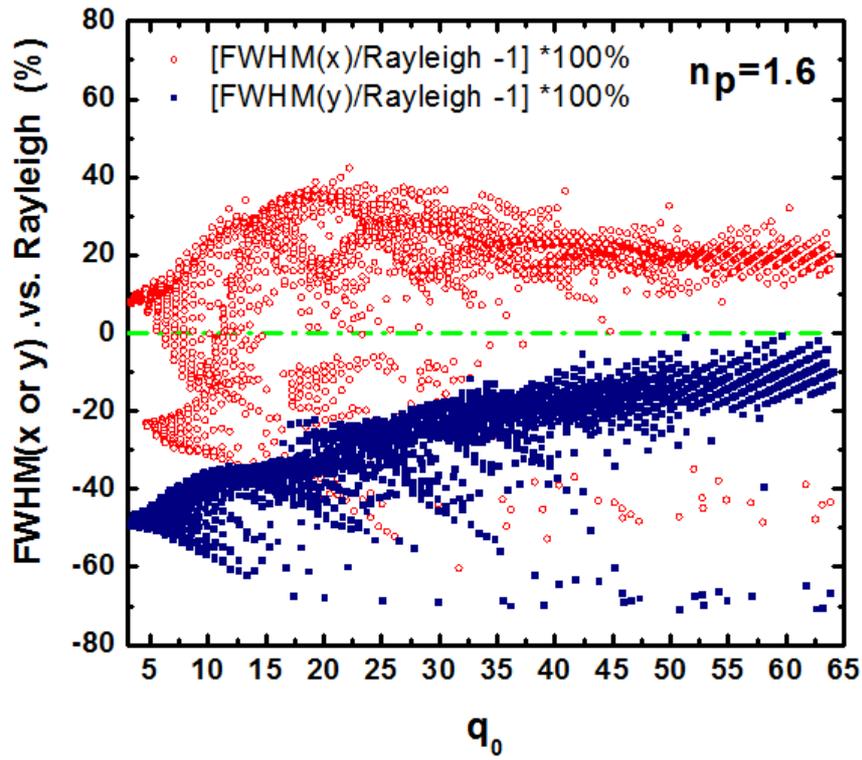

Figure 5 : FWHM(x) minus Rayleigh (square) and FWHM(y) minus Rayleigh (dot) as a function of size parameter q0. Negative values represent super-resolution in that direction.

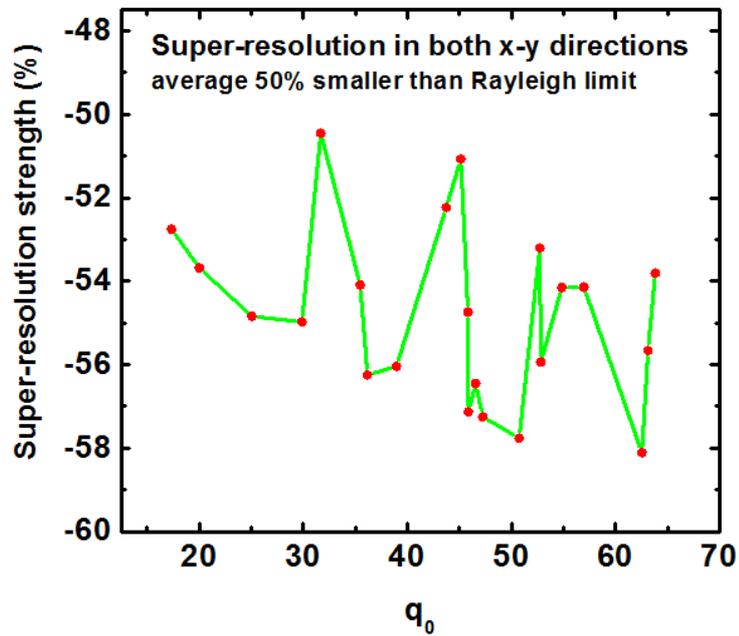

Figure 6 : Super-resolution electric field intensity ($|E|^2$) spots within size parameter range $\pi \leq q_0 \leq 20\pi$ with refractive index np=1.6 and nm=1.0.

Table 1: 'Extreme' super-resolution foci for n=1.6 microsphere

| Size parameter $q_0$ | rx= [FWHM(x)/Rayleigh-1]*100% | ry= [FWHM(y)/Rayleigh-1]*100% | Sqrt(rx*ry)*100% | Spot size in unit of $\lambda$ |
|---|---|---|---|---|
| 17.32 | -41.25356% | -67.46351% | -52.75519% | 0.288 |
| 20 | -42.59986% | -67.64719% | -53.68203% | 0.283 |
| 25.04 | -43.81465% | -68.64074% | -54.8404% | 0.275 |
| 29.86 | -43.90649% | -68.83694% | -54.97626% | 0.275 |
| 31.66 | -60.35007% | -42.17718% | -50.45192% | 0.302 |
| 35.44 | -42.67082% | -68.56142% | -54.08856% | 0.280 |
| 36.14 | -45.31019% | -69.82631% | -56.24805% | 0.267 |
| 38.92 | -45.16512% | -69.53618% | -56.04114% | 0.268 |
| 43.72 | -42.96513% | -63.49768% | -52.23204% | 0.291 |
| 45.12 | -43.49321% | -59.97436% | -51.07326% | 0.298 |
| 45.8 | -45.03153% | -66.54094% | -54.73975% | 0.276 |
| 45.84 | -47.37555% | -68.90373% | -57.13451% | 0.261 |
| 46.54 | -46.57195% | -68.42888% | -56.45233% | 0.266 |
| 47.24 | -48.23342% | -67.95402% | -57.25081% | 0.261 |
| 50.72 | -47.06663% | -70.88665% | -57.76154% | 0.258 |
| 52.7 | -42.25024% | -67.00014% | -53.205% | 0.285 |
| 52.8 | -44.89586% | -69.69272% | -55.9367% | 0.269 |
| 54.82 | -42.78771% | -68.53324% | -54.15146% | 0.280 |
| 56.94 | -43.54644% | -67.31636% | -54.14229% | 0.280 |
| <u>62.52</u> | <u>-47.80135%</u> | <u>-70.63826%</u> | <u>-58.10856%</u> | <u>0.256</u> |
| 63.1 | -44.02442% | -70.36587% | -55.65804% | 0.270 |
| 63.8 | -43.40346% | -66.70792% | -53.8085% | 0.282 |

## 4. Principles of field-invariant Mie parameter scaling

Once the super-resolution conditions were identified, it is natural to consider its scalability in different environments. For example, if the particle radius was halved, how the other Mie theory input parameters (light wavelength, background medium) should be scaled such that the final super-resolution foci remain unchanged. Such consideration is particularly important for experimental design and parameters selection for super-resolution applications. From Mie formulas, the following principles for field-invariant parameter scaling were derived:

1. $|E|^2$ field of a particle in a medium is same to that of a half-sized particle ( $a/2$ ) with doubled refractive index of particle ( $2n_p$ ) and medium ( $2n_m$ ), under a same wavelength ( $\lambda$ ) light.
2. $|E|^2$ field of a particle in a medium is same to that of a same-sized particle ( $a$ ) with doubled refractive index of particle ( $2n_p$ ) and medium ( $2n_m$ ), under a doubled wavelength ( $2\lambda$ ) light.

In Table 2, such an example parameter scaling was given. The results shown in Fig. 7 confirms that the $|E|^2$ near-field profiles are exactly the same for parameter set 1 and set 2 as in Table 2.

The above scaling are drawn based on the parameter sets listed in Table 2, with a scaling factor of 2. This scaling factor, in fact, can be simply extended to any positive number which follows Mie theory formula. Table 3 presents several scaling combinations that will produce same field profiles.

Table 2 : Input and Size Parameters

|   | Wavelength, (nm) $\lambda$ | Medium Refractive Index $n_m$ | Particle Radius (nm) $a$ | Particle Refractive Index $n_p$ | Vacuum size parameter $q_0$ | Medium size parameter $q_m$ | Particle size parameter $q_p$ |
|---|---|---|---|---|---|---|---|
| 1 | 500 | 1.0 | 500 | 1.6 | $2\pi$ | $2\pi$ | $3.2\pi$ |
| 2 | 500 | 2.0 | 250 | 3.2 | $\pi$ | $2\pi$ | $3.2\pi$ |
| 3 | 1000 | 2.0 | 500 | 3.2 | $\pi$ | $2\pi$ | $3.2\pi$ |

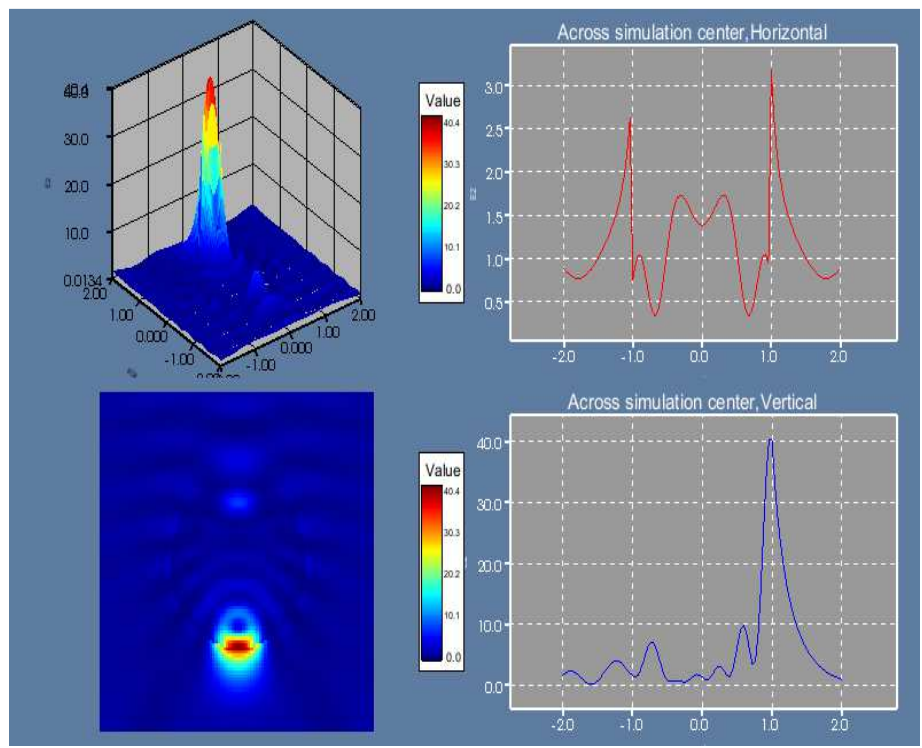

Figure 7 : $|E|^2$ field distribution for parameter set 1, 2 and 3 as in Table 2. Fields are exactly same for three cases.

Table 3: Scaling factor with Ω.

| Cases | Wavelength | Medium Refractive Index | Particle Radius | Particle Refractive Index | Notes |
|---|---|---|---|---|---|
| A | $\lambda$ | $n_m$ | $a$ | $n_p$ | Original parameter sets |
| B | $\lambda$ | $\Omega.n_m$ | $a/\Omega$ | $\Omega.n_p$ | scaling through refractive index and particle size |
| C | $\Omega.\lambda$ | $\Omega.n_m$ | $a$ | $\Omega.n_p$ | scaling through refractive index and laser wavelength |

From Table 3, one can see that, in cases B and C, the refractive index of particle has same the scaling factor as the refractive index of medium. In other words, to obtain same field focusing effect, the particle refractive index and medium refractive index must be scaled up or down **simultaneously**. As a result, we can neglect the effect of medium as it can be compensated by a particle with different refractive index. Furthermore, since $q_p = q_0.n_p = \frac{2\pi a}{\lambda}.n_p$ is correlated to $q_m = q_0.n_m = \frac{2\pi a}{\lambda}.n_m$ by $q_0 = \frac{2\pi a}{\lambda}$, we can see that independent input parameters can also be represented by ($q_0, n_p$), as mentioned above.

## 5. Conclusions:

Super-resolution and near-field scaling of a dielectric particle focusing were systematically investigated. It was shown that optical resonances inside microspheres would lead to highly oscillating features of the field focus enhancement and focus position as particle size changes. The super-resolution was observed for particular sets of size parameters and often manifests an elliptical focus shape. Extreme sharp super-resolution foci with resolution of λ/5.6 has been seen for an n=1.6 particle in air. Principles of field-invariant scaling of the Mie parameters were also derived which is important for experimental scaling of the super-resolution designs.

# References:

1. Zhang, X. and Z.W. Liu, *Superlenses to overcome the diffraction limit.* Nature Materials, 2008. **7**(6): p. 435-441.
2. Pendry, J.B., D. Schurig, and D.R. Smith, *Controlling Electromagnetic Fields.* Science, 2006. **312**: p. 1780-1782.
3. Hill, M.T., et al., *Lasing in metallic-coated nanocavities.* Nat. Photonics, 2007. **1**: p. 589.
4. Pitarke, J.M., et al., *Theory of surface plasmons and surface-plasmon polaritons.* Reports on Progress in Physics, 2007. **70**(1): p. 1-87.
5. De Leon, I. and P. Berini, *Amplification of long-range surface plasmons by a dipolar gain medium.* Nat Photon, 2010. **4**(6): p. 382-387.
6. Miroshnichenko, A.E. and Y.S. Kivshar, *Fano Resonances in All-Dielectric Oligomers.* Nano Lett., 2012. **12**(12): p. 6459-6463.
7. Guo, W., et al., *Near-field laser parallel nanofabrication of arbitrary-shaped patterns.* Appl. Phys. Lett., 2007. **90**: p. 243101.
8. Wang, Z.B., et al., *Optical near-field distribution in an asymmetrically illuminated tip-sample system for laser/STM nanopatterning.* Appl. Phys. A, 2007. **89**: p. 363-368.
9. Wang, Z., et al., *Optical virtual imaging at 50 nm lateral resolution with a white-light nanoscope.* Nat Commun, 2011. **2**: p. 218.
10. Lu, Y.F., et al., *Laser writing of sub-wavelength structure on silicon (100) surfaces with particle enhanced optical irradiation.* JEFT Letters, 2000. **72**: p. 457.
11. Wang, Z.B., et al., *A Review of Optical Near-Fields in Particle/Tip-assisted Laser Nanofabrication.* P.I.Mech. Eng. C-J. Mec., 2010. **224**: p. 1113-1127.
12. Chen, Z.G., A. Taflove, and V. Backman, *Photonic nanojet enhancement of backscattering of light by nanoparticles: a potential novel visible-light ultramicroscopy technique.* Optics Express, 2004. **12**(7): p. 1214-1220.
13. Itagi, A.V. and W.A. Challener, *Optics of photonic nanojets.* Journal of the Optical Society of America a-Optics Image Science and Vision, 2005. **22**(12): p. 2847-2858.
14. Gerlach, M., Y.P. Rakovich, and J.F. Donegan, *Nanojets and directional emission in symmetric photonic molecules.* Optics Express, 2007. **15**(25): p. 17343-17350.
15. Kong, S.C., et al., *Photonic nanojet-enabled optical data storage.* Optics Express, 2008. **16**(18): p. 13713-13719.
16. Geints, Y.E., A.A. Zemlyanov, and E.K. Panina, *Photonic nanojet effect in multilayer micrometre-sized spherical particles.* Quantum Electronics, 2011. **41**(6): p. 520-525.
17. Grojo, D., et al., *Monitoring Photonic Nanojets from Microsphere Arrays by Femtosecond Laser Ablation of Thin Films.* Journal of Nanoscience and Nanotechnology, 2011. **11**(10): p. 9129-9135.
18. Kim, M.S., et al., *Engineering photonic nanojets.* Optics Express, 2011. **19**(11): p. 10206-10220.
19. Liu, Y., C.F. Kuang, and Z.H. Ding, *Strong confinement of two-photon excitation field by photonic nanojet with radial polarization illumination.* Optics Communications, 2011.
20. Yannopapas, V., *Photonic nanojets as three-dimensional optical atom traps: A theoretical study.* Optics Communications, 2012. **285**(12): p. 2952-2955.
21. Guo, H., et al., *Near-field focusing of the dielectric microsphere with wavelength scale radius.* Opt Express, 2013. **21**(2): p. 2434-43.
22. Mie, G., *Beiträge zur Optik trüber Medien, speziell kolloidaler Metallösungen.* Ann. Phys. (Leipzig), 1908. **25**: p. 377-445.
23. Wang, Z.B., *Ph.D thesis: Optical resonance and near field effects: small particles under laser irradiation*. 2005, National University of Singapore: Singapore.